\documentclass[%
 reprint,
 amsmath,amssymb,
 aps,
 prl,
floatfix,
]{revtex4-2}

\usepackage{hyperref}
\usepackage{graphicx}
\usepackage{dcolumn}
\usepackage{bm}
\usepackage{here}
\usepackage{physics}
\usepackage{afterpage}
\hypersetup{
           breaklinks=true,   
           hidelinks                          
        }

\usepackage{ulem}
\usepackage{color}

\definecolor{violet}{rgb}{0.56, 0.0, 1.0}
\definecolor{carrotorange}{rgb}{0.93, 0.57, 0.13}

\def\eps{{\varepsilon}}

\begin{document}

\preprint{APS/123-QED}

\title{Waveform Proportionality and Taylor's Law Induced by Synchronization of Periodic and Chaotic Oscillators}

\author{Yuzuru Mitsui}
\email{mitsui@g.ecc.u-tokyo.ac.jp}
\author{Hiroshi Kori}%
 \email{kori@k.u-tokyo.ac.jp}
\affiliation{%
 Graduate School of Frontier Sciences, The University of Tokyo
}%




\date{\today}

\begin{abstract}
Taylor's law (TL), the scaling relationship between the mean and variance, has been observed in various fields. However, the underlying reasons why TL is so widely observed, why the exponents of TL are often close to 2, and the relationship between temporal and spatial TLs are not fully understood. Here, using coupled oscillator models, we analytically and numerically demonstrate that synchronization can induce TL. In particular, we show that strong synchronization leads to waveform proportionality, resulting in temporal and spatial TLs with exponent 2. Our study can help infer the existence of synchronization solely from the relationship between the mean and variance.
\end{abstract}

\maketitle

\textit{Introduction.}
Taylor's law \cite{Bliss_1941, Taylor_1961_Nature} (TL) is a power law relationship between the mean and variance:
\begin{eqnarray}
\log \mbox{(variance)} = \log \alpha + \beta \times \log \mbox{(mean)}. \label{eq:def_TL}
\end{eqnarray}
TL has been observed in various fields, such as population ecology \cite{Taylor_1961_Nature}, biophysics \cite{Sassi_etal_2022_PRX}, and complex networks \cite{Menezes_Barabasi_2004_PRL, Menezes_Barabasi_2004_PRL_2, Duch_Arenas_2006_PRL}, among others \cite{Eisler_etal_2008_AP, Taylor_2019_book}. TL is also known as fluctuation scaling in physics \cite{Eisler_etal_2008_AP}. Especially, when $\beta>1$, the relationship Eq.~(\ref{eq:def_TL}) is sometimes called giant number fluctuations. It has been investigated experimentally \cite{Narayan_etal_2007_Science} and theoretically \cite{Das_etal_2012_JTB, Houchmandzadeh_2018_PRE, Mahault_etal_2019_PRL}, and attracts significant attention in the field of active matter \cite{Toner_Tu_1995_PRL, Ginelli_2016_EPJ, Chate_2020_Annual_Review, Nishiguchi_2023_JPSJ}. TL has been extensively analyzed via theoretical studies \cite{Perry_1994_PRSB, Reuman_etal_2017_PNAS, Giometto_etal_2015_PNAS, Cohen_Xu_2015_PNAS, Sassi_etal_2022_PRX, Ballantyne_2005_EER, Ballantyne_Kerkhoff_2005_JTB, Ballantyne_Kerkhoff_2007_Oikos, Kerkhoff_Ballantyne_2003_EL, Cohen_2014_TE, Cohen_2014_TPB, Cohen_etal_2013_PRSB, Zhao_etal_2019_JAE, Cohen_2013_TPB, Cohen_Saitoh_2016_Ecology, Saitoh_2020_PE, Anderson_etal_1982_Nature, Kilpatrick_Ives_2003_Nature, Carpenter_etal_2023_TPB}, and it is usually classified into two types, i.e., temporal TL and spatial TL. In temporal TL, the means and variances are computed from data recorded at multiple time points at a given place, whereas in spatial TL, the means and variances are computed from data recorded at multiple places at a given time point.

Various studies have attempted to clarify the mechanisms of TL. Although theories show that TL exponent $\beta$ can take any real values \cite{Cohen_2014_TE, Cohen_2014_TPB, Cohen_etal_2013_PRSB}, exponents close to 2 have been often observed in real data in ecosystems for both temporal and spatial TLs \cite{Zhao_etal_2019_JAE, Taylor_Woiwod_1982_JAE, Kerkhoff_Ballantyne_2003_EL}. Cohen and Xu showed that when multiple independent random variables follow the same distribution, a correlation appears between the mean and variance by random sampling if the distribution is skewed \cite{Cohen_Xu_2015_PNAS}. While their results shed light on the ubiquity of TL, the reason for the observed TL exponents in ecosystems being often close to 2 remained unclear because TL exponents can take arbitrary values depending on the shape of distribution. By applying large deviations theory and finite-sample arguments, Giometto et al. showed that, depending on the sampling method, exponent 2 may be frequently observed in spatial TL \cite{Giometto_etal_2015_PNAS}. Reuman et al. showed that correlations between random variables affect the exponent of spatial TL \cite{Reuman_etal_2017_PNAS}. In particular, spatial TL with exponent 2 is observed when there exists a proportional relationship between time series \cite{Reuman_etal_2017_PNAS}. As a mechanism for the emergence of TL with exponents close to 2, the correlation between time series is considered crucial, and synchronization is strongly implicated as the mechanism that generates such correlations. Moreover, studies employing numerical simulations of dynamical systems have shown that synchronization affects both temporal and spatial TL exponents \cite{Cohen_Saitoh_2016_Ecology, Saitoh_2020_PE, Zhao_etal_2019_JAE}. It is noteworthy that when the degree of correlation between time series increases, exponents of temporal and spatial TLs approach 2 \cite{Ballantyne_Kerkhoff_2005_JTB, Kerkhoff_Ballantyne_2003_EL, Hanski_1987_Oikos}.

In this study, we showed that in a broad class of dynamical system models, including ecosystem models, synchronization generates a special correlation, which we call waveform proportionality, between time series, resulting in temporal and spatial TLs with exponent 2.\\
\\
\textit{Model and results.}
First, we define TL for a given time-series set $x_i(t)$ $(i=1,\ldots,N)$. For temporal TL, we compute the mean and variance of each site $i$ as ${\rm E}[x_i(t)]_t=\langle x_i(t) \rangle_t$ and ${\rm V}[x_i(t)]_t=\langle (x_i(t) - {\rm E}[x_i(t)]_t)^2 \rangle_t$, respectively, where $\langle \cdot \rangle_t$ denotes the long-time average or average over 1 cycle when $x_i(t)$ is periodic in $t$. A linear fitting to $N$ data points of $(\log {\rm E}[x_i(t)]_t, \log {\rm V}[x_i(t)]_t)$ yields slope $\beta_{\rm t}$ and intercept $\log \alpha_{\rm t}$. For spatial TL, the mean and variance at time $t$ are expressed by ${\rm E}[x_i(t)]_i = \langle x_i(t) \rangle_i$ and ${\rm V}[x_i(t)]_i=\langle (x_i(t) - {\rm E}[x_i(t)]_i)^2 \rangle_i$, respectively, where $\langle \cdot \rangle_i$ denotes the average over site $i$. A linear fitting to $M$ data points of $(\log {\rm E}[x_i(t)]_i, \log {\rm V}[x_i(t)]_i)$, where $M$ is the number of sample times, yields slope $\beta_{\rm s}$ and intercept $\log \alpha_{\rm s}$. In either case, $R^2$ denotes the coefficient of determination for linear fitting.

\begin{figure*}[htb]
\includegraphics[width=180mm]{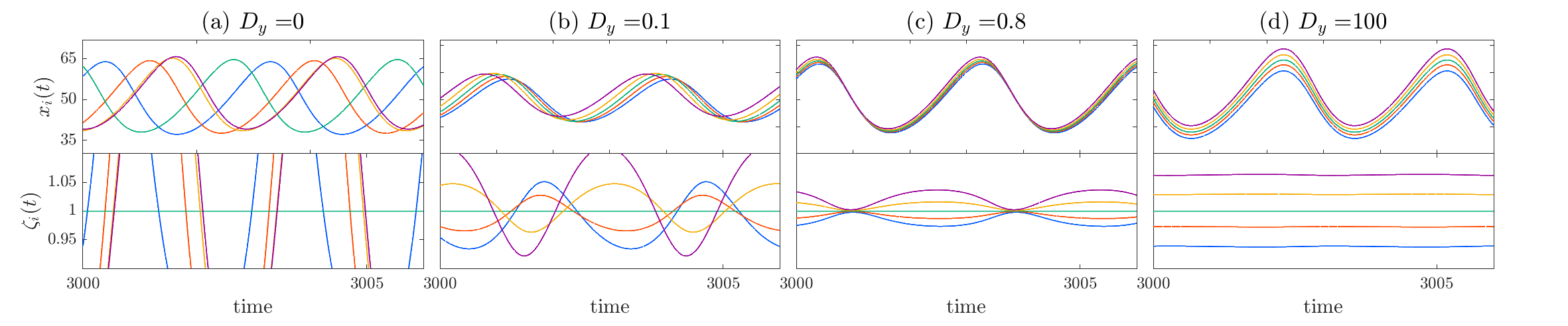}
\caption{\label{fig:food_chain_time_series_and_ratio} Examples of $x_i(t)$ and $\zeta_i(t)$ for each coupling strength. Upper and lower panels show the time series and ratio of time series, respectively. $N=100, D_x=0, D_z=0, a=1, c=9, k=0.6, l=0.1, x^*=1.6, y^*=0,$ and $z^*=0.01.$ $b_i$ was randomly selected from a uniform distribution between $4.9$ and $5.1$. Data for $i= 1,25,50,75,$ and $100$ are shown.}
\end{figure*}

Our results are based on the model describing a population of oscillators, where oscillator $i$ $(i=1,\ldots,N)$ obeys:
\begin{subequations}
\label{model}
\begin{align}
 \dot x_i &= f_x(x_i,y_i,z_i) + D_x (X-x_i), \label{model_x}\\
 \dot y_i &= f_y(x_i,y_i,z_i) + D_y (Y-y_i), \label{model_y}\\
 \dot z_i &= f_z(x_i,y_i,z_i) + D_z (Z-z_i). \label{model_z}
\end{align}
\end{subequations}
We consider a food chain model with global coupling as our first example. Concretely, we consider
$f_x = a(x_i-x^*) - lx_iy_i, f_y = -b_i(y_i-y^*)+lx_iy_i-ky_iz_i, f_z=-c(z_i-z^*) + ky_iz_i, X=\langle x_i \rangle_i, Y=\langle y_i \rangle_i,$ and $Z=\langle z_i \rangle_i$, where $x_i, y_i,$ and $z_i$ denote the populations of the vegetation, herbivores, and predators at site $i$, respectively; 
$\langle w_i \rangle_i = \frac{1}{N} \sum_{i=1}^N w_i$ is the average of population $w_i$ ($w_i=x_i, y_i, z_i$) over site $i$; and $a, b_i, c, l, k, x^*, y^*,$ and $z^*$ are parameters describing intrinsic dynamical properties \cite{Blasius_etal_1999_Nature, Blasius_Stone_2000_IJBC, Montbrio_Blasius_2003_Chaos}. Heterogeneity across sites can be expressed by parameter $b_i$ in accordance with \cite{Blasius_etal_1999_Nature, Blasius_Stone_2000_IJBC, Montbrio_Blasius_2003_Chaos}. The second terms in Eq.~\eqref{model} describe diffusive coupling with strength $D_x, D_y,$ and $D_z$. We assume that inhomogeneity can be denoted by parameter $b_i=b_0 + \mu_i$, where $b_0$ is the mean $\langle b_i \rangle_i$ and $\mu_i$ is the deviation from the mean. Note that $\langle \mu_i \rangle_i = 0$; specifically, $\mu_i$ is selected from a uniform distribution between $-0.1$ and $0.1$, and sorted in ascending order. For convenience, we introduced a reference oscillator, $i=0$, that obeys Eq.~\eqref{model} with $b_i=b_0$ and $D_x=D_y=D_z=0$. This model demonstrates synchronized oscillations for a wide range of parameters when the coupling strength is comparable or larger than ${\rm max} |\mu_i|$. The upper panels of Fig. \ref{fig:food_chain_time_series_and_ratio} illustrate the typical waveforms of $x_i(t)$. As evident in Figs. \ref{fig:food_chain_time_series_and_ratio} (b)-(d), the oscillators are synchronized in frequency in the presence of sufficiently strong coupling. 

\begin{figure}[htb]
\includegraphics[width=90mm]{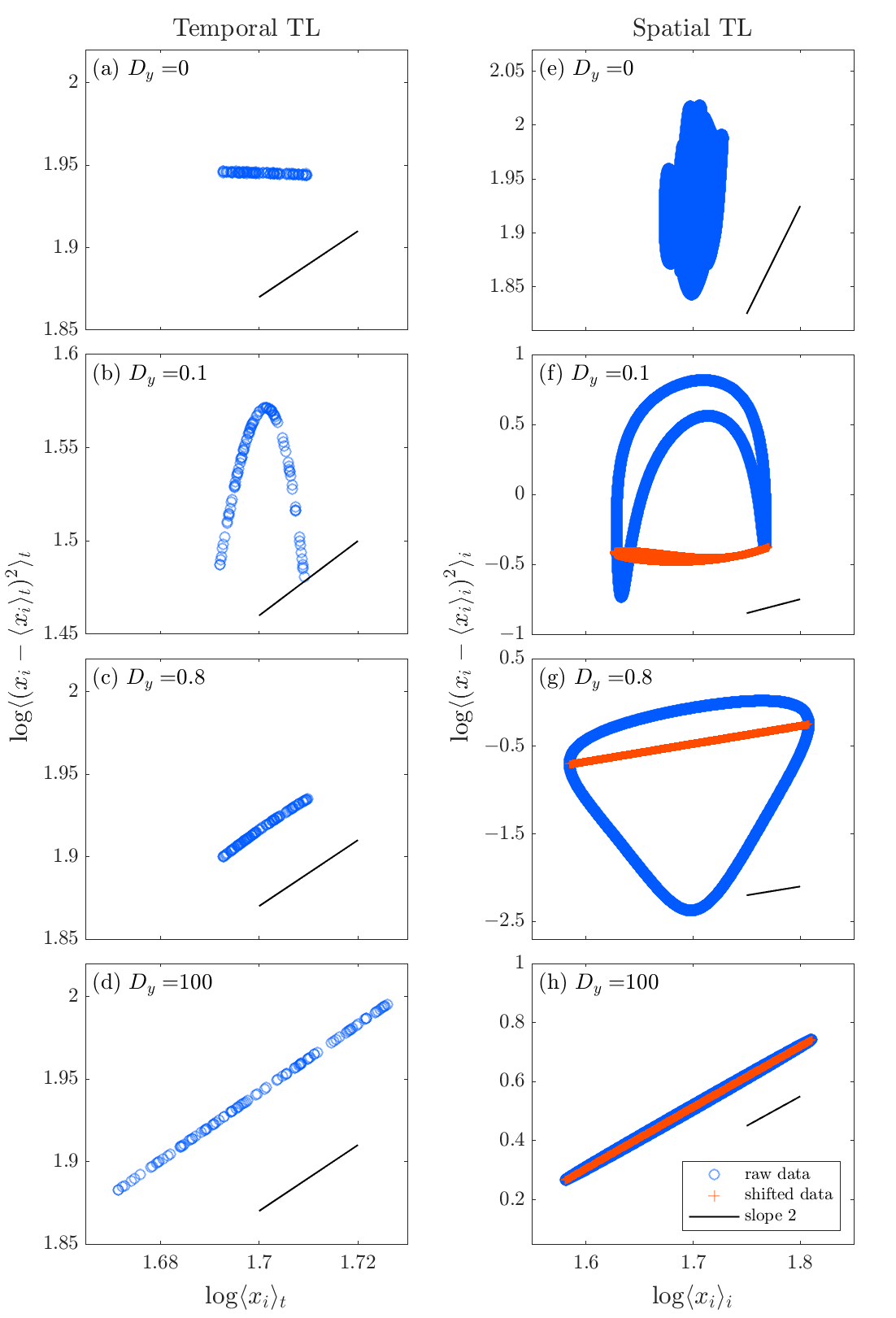}
\caption{\label{fig:food_chain_TL_example} Examples of TL for each coupling strength. $N=100, D_x=0, D_z=0, a=1, c=9, k=0.6, l=0.1, x^*=1.6, y^*=0,$ and $z^*=0.01.$ $b_i$ was randomly selected from a uniform distribution between $4.9$ and $5.1$. Temporal and spatial TLs are shown in left and right columns, respectively. Blue plots indicate temporal and spatial TLs for the raw data, red plots indicate spatial TL for the shifted data, and black lines are the reference line with slope 2.}
\end{figure}

Next, we verify temporal and spatial TLs in Fig.~\ref{fig:food_chain_TL_example}, where the blue symbols represent the mean-variance relations. TL with exponent close to 2 is observed when $D_y$ is sufficiently large. Moreover, temporal TL becomes evident for $D_y=0.8$, whereas spatial TL seems to require stronger coupling.
To determine the underlying mechanism of TL, we carefully observed the waveforms shown in the upper panel of Fig. \ref{fig:food_chain_time_series_and_ratio}(d), which yielded well-defined temporal and spatial TLs. All waveforms were found to be considerably similar. Moreover, they were approximately proportional to $x_0(t)$ (data not shown). Thus, we hypothesized that the following relation approximately holds true for all $i$ and $t$:
\begin{align}
 \label{proportionality}
 x_i(t) = C_i x_0(t),
\end{align}
where $C_i$ is constant; this relation is hereafter referred to as the {\em waveform proportionality}. Under such a relation, both temporal and spatial TLs are evidently valid with exponent $\beta_{\rm t, s}=2$, as shown below. First, note that ${\rm E}[x_i(t)]_t = C_i {\rm E}[x_0(t)]_t$, ${\rm V}[x_i(t)]_t = C_i^2 {\rm V}[x_0(t)]_t$, ${\rm E}[x_i(t)]_i = {\rm E}[C_i] x_0(t)$, and ${\rm V}[x_i(t)]_i = {\rm V}[C_i] (x_0(t))^2$. By eliminating $C_i$ and $x_0(t)$ from these relations, we obtain the following temporal and spatial TLs:
\begin{align}
 {\rm V}[x_i(t)]_t &= \alpha_{\rm t} {\rm E}[x_i(t)]_t^{\, \beta_{\rm t}}, 
 \label{temporalTL} \\
 {\rm V}[x_i(t)]_i &= \alpha_{\rm s} {\rm E}[x_i(t)]_i^{\, \beta_{\rm s}},
 \label{spatialTL}
\end{align}
where
\begin{subequations}
 \label{alpha}
 \begin{align}
 \alpha_{\rm t} &= \frac{{\rm V}[x_0(t)]_t}{{\rm E}[x_0(t)]_t^{\, 2}}, \label{alpha_t}\\
 \alpha_{\rm s} &= \frac{{\rm V}[C_i]_i}{{\rm E}[C_i]_i^{\, 2}}. \label{alpha_s}
\end{align}
\end{subequations}
Here, $\beta_{\rm t}=\beta_{\rm s}=2$. To verify this hypothesis, we plot the ratio $\zeta_i(t) = \frac{x_i(t)}{x_{j}(t)}$, where $j$ is a reference oscillator, as shown in the bottom panels of Fig.~\ref{fig:food_chain_time_series_and_ratio}. The choice of $j$ may be arbitrary; however, here, we selected $j=N/2=50$ because $x_{N/2}$ is expected to be close to $x_0$. It is now clear that waveform proportionality approximately arises in the waveforms in Fig.~\ref{fig:food_chain_time_series_and_ratio}(d) but not in those in other panels. This suggests that waveform proportionality spontaneously emerges in strongly synchronized oscillators. TL with exponent $\beta_{\rm t, s}=2$ then naturally occurs.

\begin{figure}[tb]
\includegraphics[width=90mm]{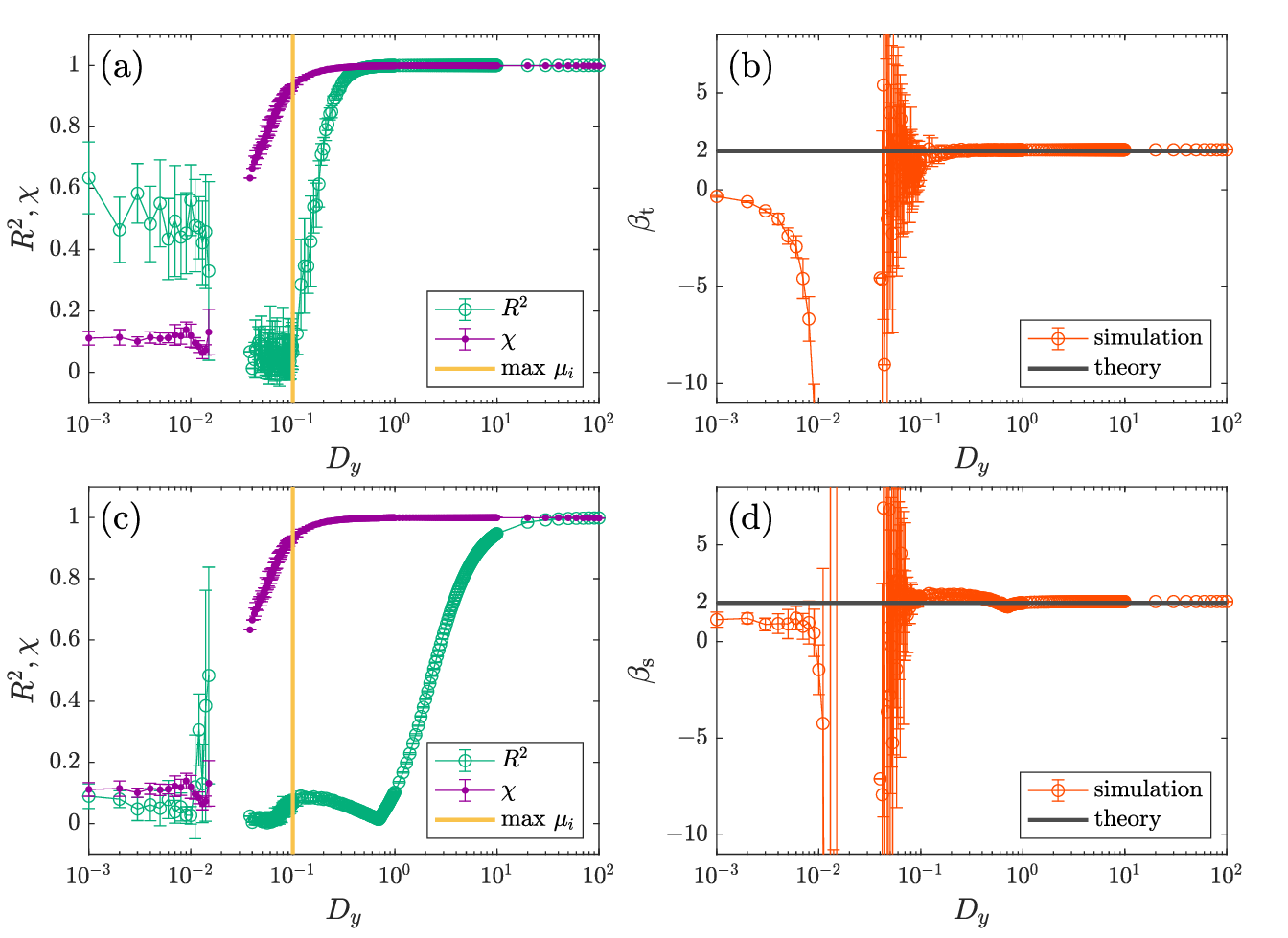}
\caption{\label{fig:food_chain_TL} $D_y$ dependence of TL parameters and synchronization degree of the food chain model. $N=100, D_x=0, D_z=0, a=1, c=9, k=0.6, l=0.1, x^*=1.6, y^*=0,$ and $z^*=0.01$. $b_i$ was randomly selected from a uniform distribution between $4.9$ and $5.1$. Simulations were performed up to $t=3500$, and TL was computed using the time series from $t=3000$ to $t=3500$. Error bars represent the standard deviation of 10 calculations with different initial conditions and $b_i$. In (a) and (c), purple and vertical orange lines represent $\chi$ and $\max\{\mu_i\}$, respectively. In (b) and (d), red and black lines represent simulation and theoretical results, respectively. (a) Coefficient of determination of temporal TL (green line). (b) Exponents of temporal TL. (c) Coefficient of determination of spatial TL (green line). (d) Exponents of spatial TL.}
\end{figure}

Next, we quantitatively investigate the dependence of the synchronization level and TL exponent on the coupling strength, $D_y$ (Fig.~\ref{fig:food_chain_TL}). Here, the order parameter $\chi$ for synchronization can be defined as 
\begin{eqnarray}
\chi = \dfrac{\mbox{CV}[X(t)]}{\underset{i}{\max}\{\mbox{CV}[x_i(t)]\}},
\end{eqnarray}
where CV represents the coefficient of variation. Namely, $\chi$ is the coefficient of variation of the mean-field of $x_i(t)$ normalized by the maximum coefficient of variation of $x_i(t)$. This quantity is close to 1 when the oscillators are completely synchronized; it is close to 0 when the oscillators are desynchronized. The same $\chi$ is plotted in Figs.~\ref{fig:food_chain_TL}(a) and (c). Moreover, $\log \alpha_{\rm t, s}$ is illustrated in Fig.~S1. Depending on the value of $D_y$, quenching may occur, rendering it impossible to define $\chi$. Thus, we judged that quenching occurred when the mean of the variance of $x_i(t)$, $1/N \sum_{i=1}^{N}\langle\left(x_i(t)-\langle x_i(t)\rangle_t\right)^2\rangle_t$, was less than a certain threshold value, and such cases were excluded. The number of times we judged that quenching occurred is shown in Fig.~S2. We confirmed that qualitatively same results could be obtained for several different threshold values. This process was applied to all systems described later. 

It can be observed in Figs.~\ref{fig:food_chain_TL}(a) and (b) that $\beta_{\rm t}$ approached a value close to 2 as $D_y$ increased. Around $D_y \approx 0.5$, $R^2$ was sufficiently close to 1 and $\beta_{\rm t}$ was close to 2. Thus, temporal TL with exponent $\beta_{\rm t} \approx 2$ was observed around $D_y = 0.5$. In contrast, according to $\chi$, synchronization began at approximately $D_y = 0.05$, significantly earlier than the onset of temporal TL. These results suggest that in addition to synchronization, there exists an unidentified condition responsible for the emergence of temporal TL. The onset of spatial TL was significantly slower than that of temporal TL. To identify the cause of this phenomenon, we focused on the waveform of $D_y=0.8$, where only temporal TL was observed (Figs.~\ref{fig:food_chain_TL_example}(c) and (g)). Here, waveform proportionality was not realized well (Fig.~\ref{fig:food_chain_time_series_and_ratio}(c), bottom); however, when the waveforms were shifted such that their peak positions coincided, the ratio $\zeta_i(t)$ became almost constant, as shown in Fig.~S3(c). This indicates that the hypothesis considered in Eq.~\eqref{proportionality} should be replaced with $x_i(t) = C_i x_0(t - t_i)$. We further found the shift $t_i$ to be approximately proportional to $\mu_i$ and $1/D_y$ (Figs.~S4 and S5). These results suggest that waveform proportionality occurs with a lag $t_i$, which decreases when $D_y$ is increased. Because Eq.~\eqref{temporalTL} holds true even when $x_i(t)= C_i x_0(t)$ is replaced with $x_i(t)= C_i x_0(t-t_i)$, temporal TL emerges at smaller $D_y$. In contrast, Eq.~\eqref{spatialTL} is violated in the presence of lag; thus, spatial TL may appear only when the lag is vanishingly small, i.e., $D_y$ is considerably large. Accordingly, we found that spatial TL is observed for a wider range of $D_y$ upon using shifted waveforms $x_i(t+t_i)$, as shown in red symbols in Figs.~\ref{fig:food_chain_TL_example}(g) and (h), and Fig.~S6.

To theoretically clarify the mechanism responsible for the emergence of TL, we performed a perturbative analysis. Motivated by our numerical results, we considered the following ansatzes:
\begin{subequations}
  \label{ansatz}
\begin{eqnarray}
 x_i(t) &=& x_0(t- \eps_i \tau) + \eps_i p(t- \eps_i \tau) + O(\eps_i^2), \label{x_ansatz}\\
 y_i(t) &=& y_0(t- \eps_i \tau) + \eps_i q(t- \eps_i \tau) + O(\eps_i^2), \label{y_ansatz}\\
 z_i(t) &=& z_0(t- \eps_i \tau) + \eps_i r(t- \eps_i \tau) + O(\eps_i^2), \label{z_ansatz}
\end{eqnarray}
\end{subequations}
where
\begin{equation}
 \eps_i = \frac{\mu_i}{D_y},
\end{equation}
denotes nondimensional small parameters; $\tau$ is a constant; and $p(t), q(t)$, and $r(t)$ are functions to be determined. We recall that $x_0(t)$ is a periodic function obeying the following relation for the food chain model under consideration:
\begin{equation}
 \dot{x}_0 = f_x(x_0,y_0,z_0) = (a - l y_0) x_0- a x^*. \label{dot_x0}
\end{equation}
In Eq.~\eqref{ansatz}, waveform proportionality can be said to occur in variable $x_i(t)$ if $p(t) \propto x_0(t)$ holds true in good approximation. As shown below, this is true under some conditions. Substituting Eq.~\eqref{ansatz} into Eq.~\eqref{model_x} and extracting the $O(\varepsilon_i)$ terms, we obtain:
\begin{equation}
\dot{p} = (a - ly_0-D_x)p -lqx_0. \label{dot_p}
\end{equation}
Notice that Eqs.~\eqref{dot_x0} and \eqref{dot_p} are linear in terms of $x_0$ and $p$, respectively. Therefore, by assuming that other time-dependent functions are provided, we may solve these equations to obtain the expressions for periodic $x_0(t)$ and $p(t)$. As shown in Supplemental Material, we obtain:
\begin{subequations}
\begin{align}
 x_0(t) &= \left(\kappa_1-ax^* \int_0^t e^{-\bar f t'} g(t')dt'\right)e^{\bar f t + \delta F(t)}, \label{x0} \\
 p(t) &= \left(\kappa_2-l\int_0^t  e^{-(\bar f -D_x) t'} h(t') dt'\right)e^{(\bar f -D_x)t + \delta F(t)},
 \label{p}
\end{align}
\end{subequations}
where $\bar f = \langle a - ly_0(t) \rangle_t, \delta F(t) = \int_{0}^{t} \left\{f(t') - \bar f \right\}dt', g(t) =  e^{- \delta F(t)},$ and $h(t) =  q(t) x_0(t) e^{-\delta F(t)}$. $\kappa_1$ and $\kappa_2$ are arbitrary constants. Note that $\delta F(t)$ is periodic, and thus, $g(t)$ and $h(t)$ are periodic. In the integral provided in Eq.~\eqref{x0}, if the oscillation period of $g(t)$, which is approximately $2 \pi/ \omega$, is sufficiently smaller than $1/|\bar f|$, $e^{-\bar f t'}$ is approximately constant during 1 oscillation period, where $\omega$ is the oscillation frequency. Then, $g(t)$ can be effectively averaged, and only the average of $g(t)$ is important. By using the Fourier series of $g(t)$, the integral can be directly calculated to obtain:
\begin{equation}
   x_0 = \frac{a x^*}{\bar f} \left[A+ O\left( \frac{\bar f}{\omega}\right) \right] e^{\delta F(t)}, \label{x0_expression}
\end{equation}
where $A=\langle g(t) \rangle_t$. Similarly, we obtain:
\begin{align}
 p = \frac{l}{\bar f-D_x} \left[
B+ O\left( \frac{\bar f-D_x}{\omega}\right) \right] e^{\delta F(t)}, \label{p_expression}
\end{align}
where $B = \langle q(t) x_0(t) e^{-\delta F(t)} \rangle_t$. Therefore, $p(t)$ becomes approximately proportional to $x_0(t)$ when the following conditions are satisfied:
\begin{align}
 \label{condition}
 A \gg O\left( \frac{\bar f}{\omega}\right) 
\mbox{ and } B \gg O\left( \frac{\bar f-D_x}{\omega}\right).
\end{align}
Thus, TL with exponent $\beta_{\rm t,s}=2$ should be observed in good approximation when $\eps_i$ is sufficiently small. Furthermore, substituting Eqs.~\eqref{x0_expression} and \eqref{p_expression} into Eqs.~\eqref{temporalTL} and \eqref{spatialTL}, and omitting the $O(\cdot)$ terms, we obtain: 
\begin{align}
 \alpha_{\rm t} &= \dfrac{\mbox{V}\left[e^{\delta F(t)}\right]_t}{\left(\mbox{E}\left[e^{\delta F(t)}\right]_t\right)^2},\\
 \alpha_{\rm s} &= \left(\frac{\bar f}{\bar f-D_x}\frac{l B}{a x^* A}\right)^2 {\rm V}[\eps_i]_i.
\end{align}
We expect that Eq.~\eqref{condition} can be generally satisfied when $\bar f$ and $D_x$ are sufficiently smaller than $\omega$. In the present example, we have $\omega \approx 2.18, \bar f \approx 0.0327,$ and $D_x=0$, suggesting the validity of our approximation. 
Indeed, the predicted $\beta_{\rm t,s}$ and $\log \alpha_{\rm t,s}$, shown in black lines in Figs.~\ref{fig:food_chain_TL} and S1, are in excellent agreement with the simulation results for large $D_y$.
One might naively expect that waveform proportionality naturally arises for oscillators with strong diffusive coupling because the waveforms become virtually identical in the strong coupling limit. However, note that convergence ($p(t) \to 0$) does not imply waveform proportionality, and the convergence method is important. An interesting prediction, possibly opposing the naive expectation, is that waveform proportionality is violated when coupling strength $D_x$ in the observed variable is large because this will violate Eq.~\eqref{condition}. We numerically demonstrate this prediction in Supplemental Material by considering the $D_x>0$ cases (Fig.~S7). In contrast, there is no condition regarding $D_z$.  Indeed, as demonstrated in Fig.~S8, TL is observed for $D_z>0$.

\begin{figure}[htb]
\includegraphics[width=90mm]{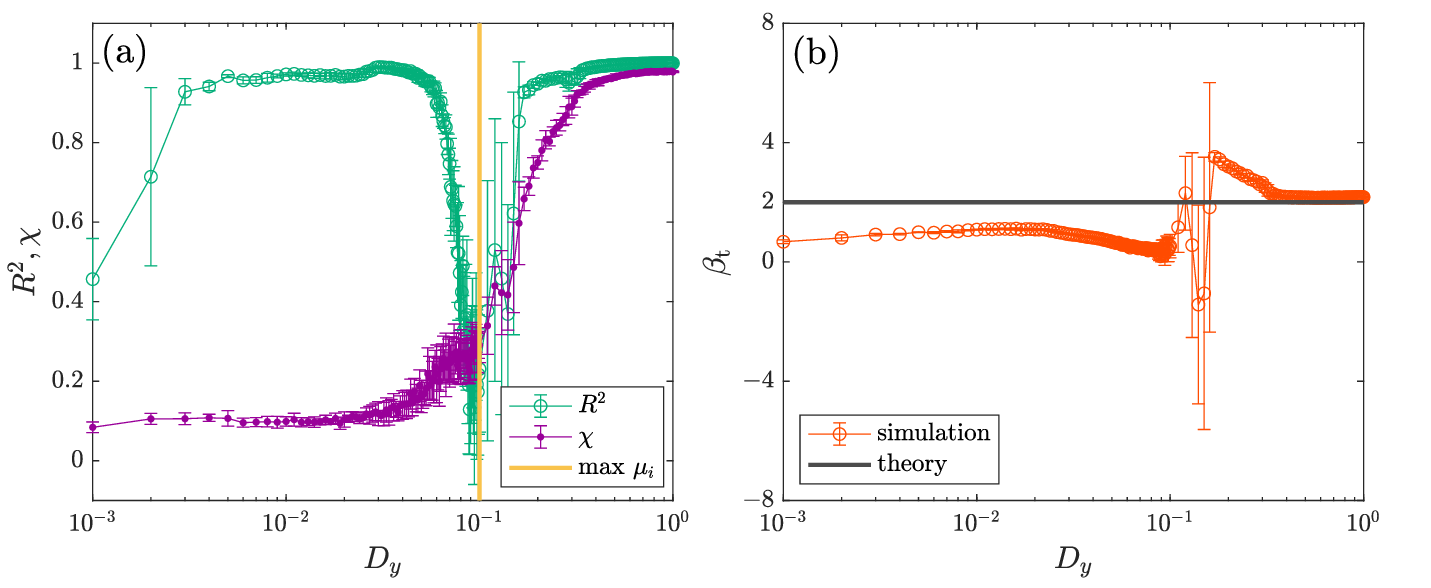}
\caption{\label{fig:chaotic_food_chain_TL} $D_y$ dependence of TL parameters and synchronization degree of the food chain model in chaotic region. $N=100, a=1, c=10, k=0.6, l=0.1, x^*=1.5, y^*=0,$ and $z^*=0.01.$ $b_i$ was randomly selected from a uniform distribution between $0.9$ and $1.1$. Simulations were performed up to $t=3500$, and TL was computed using the time series from $t=3000$ to $t=3500$. Error bars represent the standard deviation of 10 calculations with different initial conditions and $b_i$. In (a), purple and vertical orange lines represent $\chi$ and $\max\{\mu_i\}$, respectively. In (b), red and black lines represent simulation and theoretical results, respectively. (a) Coefficient of determination of temporal TL (green line). (b) Exponents of temporal TL.}
\end{figure}

Next, we generalize our theory considering the following situation. Suppose we have $N$ oscillators, each of which can be described by an $M$-dimensional dynamical system. Let $x_i(t)$ $(i=1,\ldots,N)$ be the observables obeying
\begin{equation}
\dot x_i = s_i(t) x_0 + u_i(t),
\label{general}
\end{equation}
where $s_i(t)$ and $u_i(t)$ are periodic with period $\frac{2\pi}{\omega}$. We assume that $s_i(t) = s(t) + \eps_i \delta s(t)$ and $u_i(t)= u(t) + \eps_i \delta u(t)$ to the lowest order in $\eps_i$, where $s, u, \delta s,$ and $\delta u$ are periodic. Then, the above analysis can similarly be applied to this system. Accordingly, we conclude that waveform proportionality occurs in $x_i(t)$ for small $\eps_i$ if Eq.~\eqref{condition}, wherein $\bar f$ is replaced with $\langle s(t) \rangle_t$, is satisfied. Essentially, the equation should be linear in terms of observables and its intrinsic dynamics should be sufficiently slow. When these assumptions are satisfied, the averaging approximation is effective, resulting in waveform proportionality. Furthermore, we note that our theory can approximately be extended to a class of chaotic oscillators. Suppose that $s_i(t)$ and $u_i(t)$ in Eq.~\eqref{general} show chaotic oscillations with characteristic period $T$. We assume that the time averages of $s_i(t)$ and $u_i(t)$ over 1 period $T$ do not strongly fluctuate from the long-time averages of $s_i(t)$ and $u_i(t)$. Under this assumption, the above arguments hold approximately true.  From these observations, we predict that TL can arise in a broad class of systems, including various types of (i) chaotic oscillators, (ii) coupling mechanisms, and (iii) dynamical systems. Concerning (i), as shown in Fig.~\ref{fig:chaotic_food_chain_TL} and Fig.~S9, we demonstrate that the same food chain model with another set of parameters that yields chaotic oscillations approximately shows waveform proportionality and TL when the oscillators are synchronized. Concerning (ii), as shown in Fig.~S10, we demonstrate that TL is observed in a pacemaker-driven system, where $Y = y_0$ in Eq.~\eqref{model}. Concerning (iii), as an example, we consider the following R\"{o}ssler system \cite{Rosenblum_etal_1996_PRL, Rosenblum_etal_1997_PRL, Sakaguchi_2000_PRE, Pikovsky_etal_1996_EPL,Montbrio_Blasius_2003_Chaos, Osipov_etal_1997_PRE}: $f_x= -(\omega_0+\mu_i) y_i - z_i, f_y=(\omega_0+\mu_i) x_i + ay_i,$ and $f_z=b + z_i\left(x_i - c\right)$ in Eq.~\eqref{model}, where we employ the standard parameter values as $a=0.1, b=0.1,$ and $c=0.7$, and introduce $\mu_i$ as a heterogeneity parameter. Following previous works \cite{Rosenblum_etal_1996_PRL, Rosenblum_etal_1997_PRL, Sakaguchi_2000_PRE, Pikovsky_etal_1996_EPL,Montbrio_Blasius_2003_Chaos, Osipov_etal_1997_PRE}, we consider $D_x=D_y=D>0$ and $D_z=0$. The simulation results of the R\"{o}ssler system are shown in Fig.~S11. Note that the actual frequency, $\omega$, is approximately $\omega_0$ in this system. In this system, TL is expected to be observed in variable $z$ because Eq.~\eqref{condition}, wherein $D_x$ and $\bar f$ are replaced with $0$ and $\langle x_i(t) - c \rangle_t$, respectively, may be satisfied for any value of $D$. Furthermore, we investigate $\omega_0$ dependency in addition to $D$ dependency because the validity of Eq.~\eqref{condition} can conveniently be controlled through $\omega_0$. We investigate $\omega_0$ dependency for fixed $D=100$. As expected,
exponent $\beta_{\rm t,s}$ approaches $2$ with large coefficients of determination ($R^2$) as $\omega_0$ increases (Fig.~S12). 
\\
\\
\textit{Discussion.}
In this study, we showed that temporal and spatial TLs are induced by synchronization in a broad class of periodic and chaotic oscillators. Specifically, we demonstrated that as the degree of synchronization increases, the correlation between log(mean) and log(variance) becomes stronger. Moreover, we showed that in regions of strong synchronization, waveform proportionality emerges, resulting in the derivation of temporal and spatial TLs with exponent 2. In these synchronization-induced TLs, temporal and spatial TLs arise in the same mechanism, i.e., waveform proportionality. In contrast, owing to phase lag, which is expressed by the $\varepsilon_i \tau$ terms in Eq.~\eqref{ansatz}, spatial TL requires stronger coupling than temporal TL. While several studies explored the relationship between TL and synchronization \cite{Reuman_etal_2017_PNAS, Cohen_Saitoh_2016_Ecology, Saitoh_2020_PE, Ballantyne_Kerkhoff_2005_JTB, Ballantyne_Kerkhoff_2007_Oikos, Hanski_1987_Oikos, Kerkhoff_Ballantyne_2003_EL, Hanski_Woiwod_1993_JAE, Eisler_etal_2008_AP, Zhao_etal_2019_JAE}, the details of the relationship are not fully understood yet. Many of these studies investigated the correlation between the exponents of TL and degree of synchronization through numerical simulations \cite{Cohen_Saitoh_2016_Ecology, Saitoh_2020_PE, Ballantyne_Kerkhoff_2005_JTB, Hanski_1987_Oikos, Kerkhoff_Ballantyne_2003_EL, Zhao_etal_2019_JAE}. In refs. \cite{Eisler_etal_2008_AP, Ballantyne_Kerkhoff_2005_JTB, Ballantyne_Kerkhoff_2007_Oikos}, the authors argued that temporal TL with exponent 2 is observed when the time series are perfectly correlated. Reuman et al. derived the analytical relationship between spatial TL and synchronization \cite{Reuman_etal_2017_PNAS}. In this study, we analytically derived both temporal and spatial TLs with exponent 2 from the synchronization state in a broad class of periodic and chaotic oscillators. Although other mechanisms are known to exist for TL with exponent 2 \cite{Ballantyne_2005_EER, Cohen_2014_TE, Cohen_etal_2013_PRSB, Sassi_etal_2022_PRX, Cohen_2013_TPB, Giometto_etal_2015_PNAS, Kilpatrick_Ives_2003_Nature, Anderson_etal_1982_Nature, Carpenter_etal_2023_TPB}, and the synchronization considered in this study is slightly unusual owing to its intensity and special correlation (waveform proportionality), we believe that our findings provide valuable insights into the understanding of TL from the perspective of synchronization as another universal phenomenon in ecosystems \cite{Liebhold_etal_2004_review}. Our study can help in inferring the existence of synchronization solely from the relationship between the mean and variance \cite{Petri_etal_2013_SR}.

Similar phenomena as waveform proportionality have been proposed previously, including the projective synchronization \cite{Mainieri_Rehacek_1999_PRL}
and a type of generalized synchronization \cite{Kano_Umeno_2022_Chaos}.
These previously reported phenomena are observed in a limited class of coupled oscillators. We showed that waveform proportionality arises in a broad class of coupled periodic or chaotic oscillators when the oscillators are strongly synchronized.

This study was supported by JSPS KAKENHI Grant Number JP23KJ0830 and WINGS CFS program of The Univ. of Tokyo to Y.M., and JSPS KAKENHI Grant Number JP21K12056 to H.K.



\bibliography{reference}

\begin{thebibliography}{49}%
\makeatletter
\providecommand \@ifxundefined [1]{%
 \@ifx{#1\undefined}
}%
\providecommand \@ifnum [1]{%
 \ifnum #1\expandafter \@firstoftwo
 \else \expandafter \@secondoftwo
 \fi
}%
\providecommand \@ifx [1]{%
 \ifx #1\expandafter \@firstoftwo
 \else \expandafter \@secondoftwo
 \fi
}%
\providecommand \natexlab [1]{#1}%
\providecommand \enquote  [1]{``#1''}%
\providecommand \bibnamefont  [1]{#1}%
\providecommand \bibfnamefont [1]{#1}%
\providecommand \citenamefont [1]{#1}%
\providecommand \href@noop [0]{\@secondoftwo}%
\providecommand \href [0]{\begingroup \@sanitize@url \@href}%
\providecommand \@href[1]{\@@startlink{#1}\@@href}%
\providecommand \@@href[1]{\endgroup#1\@@endlink}%
\providecommand \@sanitize@url [0]{\catcode `\\12\catcode `\$12\catcode
  `\&12\catcode `\#12\catcode `\^12\catcode `\_12\catcode `\%12\relax}%
\providecommand \@@startlink[1]{}%
\providecommand \@@endlink[0]{}%
\providecommand \url  [0]{\begingroup\@sanitize@url \@url }%
\providecommand \@url [1]{\endgroup\@href {#1}{\urlprefix }}%
\providecommand \urlprefix  [0]{URL }%
\providecommand \Eprint [0]{\href }%
\providecommand \doibase [0]{https://doi.org/}%
\providecommand \selectlanguage [0]{\@gobble}%
\providecommand \bibinfo  [0]{\@secondoftwo}%
\providecommand \bibfield  [0]{\@secondoftwo}%
\providecommand \translation [1]{[#1]}%
\providecommand \BibitemOpen [0]{}%
\providecommand \bibitemStop [0]{}%
\providecommand \bibitemNoStop [0]{.\EOS\space}%
\providecommand \EOS [0]{\spacefactor3000\relax}%
\providecommand \BibitemShut  [1]{\csname bibitem#1\endcsname}%
\let\auto@bib@innerbib\@empty
\bibitem [{\citenamefont {Bliss}(1941)}]{Bliss_1941}%
  \BibitemOpen
  \bibfield  {author} {\bibinfo {author} {\bibfnamefont {C.~I.}\ \bibnamefont
  {Bliss}},\ }\href {https://doi.org/10.1093/jee/34.2.221} {\bibfield
  {journal} {\bibinfo  {journal} {J. Econ. Entomol.}\ }\textbf {\bibinfo
  {volume} {34}},\ \bibinfo {pages} {221} (\bibinfo {year} {1941})}\BibitemShut
  {NoStop}%
\bibitem [{\citenamefont {Taylor}(1961)}]{Taylor_1961_Nature}%
  \BibitemOpen
  \bibfield  {author} {\bibinfo {author} {\bibfnamefont {L.~R.}\ \bibnamefont
  {Taylor}},\ }\href {https://doi.org/10.1038/189732a0} {\bibfield  {journal}
  {\bibinfo  {journal} {Nature}\ }\textbf {\bibinfo {volume} {189}},\ \bibinfo
  {pages} {732} (\bibinfo {year} {1961})}\BibitemShut {NoStop}%
\bibitem [{\citenamefont {Sassi}\ \emph {et~al.}(2022)\citenamefont {Sassi},
  \citenamefont {Garcia-Alcala}, \citenamefont {Aldana},\ and\ \citenamefont
  {Tu}}]{Sassi_etal_2022_PRX}%
  \BibitemOpen
  \bibfield  {author} {\bibinfo {author} {\bibfnamefont {A.~S.}\ \bibnamefont
  {Sassi}}, \bibinfo {author} {\bibfnamefont {M.}~\bibnamefont
  {Garcia-Alcala}}, \bibinfo {author} {\bibfnamefont {M.}~\bibnamefont
  {Aldana}},\ and\ \bibinfo {author} {\bibfnamefont {Y.}~\bibnamefont {Tu}},\
  }\href {https://doi.org/10.1103/PhysRevX.12.011051} {\bibfield  {journal}
  {\bibinfo  {journal} {Phys. Rev. X.}\ }\textbf {\bibinfo {volume} {12}},\
  \bibinfo {pages} {011051} (\bibinfo {year} {2022})}\BibitemShut {NoStop}%
\bibitem [{\citenamefont {de~Menezes}\ and\ \citenamefont
  {Barab^^c3^^a1si}(2004)}]{Menezes_Barabasi_2004_PRL}%
  \BibitemOpen
  \bibfield  {author} {\bibinfo {author} {\bibfnamefont {M.~A.}\ \bibnamefont
  {de~Menezes}}\ and\ \bibinfo {author} {\bibfnamefont {A.~L.}\ \bibnamefont
  {Barab^^c3^^a1si}},\ }\href {https://doi.org/10.1103/PhysRevLett.92.028701}
  {\bibfield  {journal} {\bibinfo  {journal} {Phys. Rev. Lett.}\ }\textbf
  {\bibinfo {volume} {92}},\ \bibinfo {pages} {028701} (\bibinfo {year}
  {2004})}\BibitemShut {NoStop}%
\bibitem [{\citenamefont {Menezes}\ and\ \citenamefont
  {Barab^^c3^^a1si}(2004)}]{Menezes_Barabasi_2004_PRL_2}%
  \BibitemOpen
  \bibfield  {author} {\bibinfo {author} {\bibfnamefont {M.~A.~D.}\
  \bibnamefont {Menezes}}\ and\ \bibinfo {author} {\bibfnamefont {A.~L.}\
  \bibnamefont {Barab^^c3^^a1si}},\ }\href
  {https://doi.org/10.1103/PhysRevLett.93.068701} {\bibfield  {journal}
  {\bibinfo  {journal} {Phys. Rev. Lett.}\ }\textbf {\bibinfo {volume} {93}},\
  \bibinfo {pages} {068701} (\bibinfo {year} {2004})}\BibitemShut {NoStop}%
\bibitem [{\citenamefont {Duch}\ and\ \citenamefont
  {Arenas}(2006)}]{Duch_Arenas_2006_PRL}%
  \BibitemOpen
  \bibfield  {author} {\bibinfo {author} {\bibfnamefont {J.}~\bibnamefont
  {Duch}}\ and\ \bibinfo {author} {\bibfnamefont {A.}~\bibnamefont {Arenas}},\
  }\href {https://doi.org/10.1103/PhysRevLett.96.218702} {\bibfield  {journal}
  {\bibinfo  {journal} {Phys. Rev. Lett.}\ }\textbf {\bibinfo {volume} {96}},\
  \bibinfo {pages} {218702} (\bibinfo {year} {2006})}\BibitemShut {NoStop}%
\bibitem [{\citenamefont {Eisler}\ \emph {et~al.}(2008)\citenamefont {Eisler},
  \citenamefont {Bartos},\ and\ \citenamefont
  {Kert{\'{e}}sz}}]{Eisler_etal_2008_AP}%
  \BibitemOpen
  \bibfield  {author} {\bibinfo {author} {\bibfnamefont {Z.}~\bibnamefont
  {Eisler}}, \bibinfo {author} {\bibfnamefont {I.}~\bibnamefont {Bartos}},\
  and\ \bibinfo {author} {\bibfnamefont {J.}~\bibnamefont {Kert{\'{e}}sz}},\
  }\href {https://doi.org/10.1080/00018730801893043} {\bibfield  {journal}
  {\bibinfo  {journal} {Adv. Phys.}\ }\textbf {\bibinfo {volume} {57}},\
  \bibinfo {pages} {89} (\bibinfo {year} {2008})}\BibitemShut {NoStop}%
\bibitem [{\citenamefont {Taylor}(2019)}]{Taylor_2019_book}%
  \BibitemOpen
  \bibfield  {author} {\bibinfo {author} {\bibfnamefont {R.~A.~J.}\
  \bibnamefont {Taylor}},\ }\href@noop {} {\emph {\bibinfo {title} {{Taylor's
  Power Law: Order and Pattern in Nature}}}},\ \bibinfo {edition} {1st}\ ed.\
  (\bibinfo  {publisher} {Academic Press},\ \bibinfo {year} {2019})\BibitemShut
  {NoStop}%
\bibitem [{\citenamefont {Narayan}\ \emph {et~al.}(2007)\citenamefont
  {Narayan}, \citenamefont {Ramaswamy},\ and\ \citenamefont
  {Menon}}]{Narayan_etal_2007_Science}%
  \BibitemOpen
  \bibfield  {author} {\bibinfo {author} {\bibfnamefont {V.}~\bibnamefont
  {Narayan}}, \bibinfo {author} {\bibfnamefont {S.}~\bibnamefont {Ramaswamy}},\
  and\ \bibinfo {author} {\bibfnamefont {N.}~\bibnamefont {Menon}},\ }\href
  {https://doi.org/10.1126/science.1142373} {\bibfield  {journal} {\bibinfo
  {journal} {Science}\ }\textbf {\bibinfo {volume} {317}},\ \bibinfo {pages}
  {105} (\bibinfo {year} {2007})}\BibitemShut {NoStop}%
\bibitem [{\citenamefont {Das}\ \emph {et~al.}(2012)\citenamefont {Das},
  \citenamefont {Das},\ and\ \citenamefont {Prasad}}]{Das_etal_2012_JTB}%
  \BibitemOpen
  \bibfield  {author} {\bibinfo {author} {\bibfnamefont {D.}~\bibnamefont
  {Das}}, \bibinfo {author} {\bibfnamefont {D.}~\bibnamefont {Das}},\ and\
  \bibinfo {author} {\bibfnamefont {A.}~\bibnamefont {Prasad}},\ }\href
  {https://doi.org/10.1016/j.jtbi.2012.05.030} {\bibfield  {journal} {\bibinfo
  {journal} {J. Theor. Biol.}\ }\textbf {\bibinfo {volume} {308}},\ \bibinfo
  {pages} {96} (\bibinfo {year} {2012})}\BibitemShut {NoStop}%
\bibitem [{\citenamefont {Houchmandzadeh}(2018)}]{Houchmandzadeh_2018_PRE}%
  \BibitemOpen
  \bibfield  {author} {\bibinfo {author} {\bibfnamefont {B.}~\bibnamefont
  {Houchmandzadeh}},\ }\href@noop {} {\bibfield  {journal} {\bibinfo  {journal}
  {Phys. Rev. E}\ }\textbf {\bibinfo {volume} {98}},\ \bibinfo {pages} {042118}
  (\bibinfo {year} {2018})}\BibitemShut {NoStop}%
\bibitem [{\citenamefont {Mahault}\ \emph {et~al.}(2019)\citenamefont
  {Mahault}, \citenamefont {Ginelli},\ and\ \citenamefont
  {Chat^^c3^^a9}}]{Mahault_etal_2019_PRL}%
  \BibitemOpen
  \bibfield  {author} {\bibinfo {author} {\bibfnamefont {B.}~\bibnamefont
  {Mahault}}, \bibinfo {author} {\bibfnamefont {F.}~\bibnamefont {Ginelli}},\
  and\ \bibinfo {author} {\bibfnamefont {H.}~\bibnamefont {Chat^^c3^^a9}},\
  }\href@noop {} {\bibfield  {journal} {\bibinfo  {journal} {Phys. Rev. Lett.}\
  }\textbf {\bibinfo {volume} {123}},\ \bibinfo {pages} {218001} (\bibinfo
  {year} {2019})}\BibitemShut {NoStop}%
\bibitem [{\citenamefont {Toner}\ and\ \citenamefont
  {Tu}(1995)}]{Toner_Tu_1995_PRL}%
  \BibitemOpen
  \bibfield  {author} {\bibinfo {author} {\bibfnamefont {J.}~\bibnamefont
  {Toner}}\ and\ \bibinfo {author} {\bibfnamefont {Y.}~\bibnamefont {Tu}},\
  }\href@noop {} {\bibfield  {journal} {\bibinfo  {journal} {Phys. Rev. Lett.}\
  }\textbf {\bibinfo {volume} {75}},\ \bibinfo {pages} {4326} (\bibinfo {year}
  {1995})}\BibitemShut {NoStop}%
\bibitem [{\citenamefont {Ginelli}(2016)}]{Ginelli_2016_EPJ}%
  \BibitemOpen
  \bibfield  {author} {\bibinfo {author} {\bibfnamefont {F.}~\bibnamefont
  {Ginelli}},\ }\href {https://doi.org/10.1140/epjst/e2016-60066-8} {\bibfield
  {journal} {\bibinfo  {journal} {Eur. Phys. J. Special Topics}\ }\textbf
  {\bibinfo {volume} {225}},\ \bibinfo {pages} {2099} (\bibinfo {year}
  {2016})}\BibitemShut {NoStop}%
\bibitem [{\citenamefont {Chat^^c3^^a9}(2020)}]{Chate_2020_Annual_Review}%
  \BibitemOpen
  \bibfield  {author} {\bibinfo {author} {\bibfnamefont {H.}~\bibnamefont
  {Chat^^c3^^a9}},\ }\href {https://doi.org/10.1146/annurev-conmatphys}
  {\bibfield  {journal} {\bibinfo  {journal} {Annu. Rev. Condens. Matter
  Phys.}\ }\textbf {\bibinfo {volume} {11}},\ \bibinfo {pages} {189} (\bibinfo
  {year} {2020})}\BibitemShut {NoStop}%
\bibitem [{\citenamefont {Nishiguchi}(2023)}]{Nishiguchi_2023_JPSJ}%
  \BibitemOpen
  \bibfield  {author} {\bibinfo {author} {\bibfnamefont {D.}~\bibnamefont
  {Nishiguchi}},\ }\href@noop {} {\bibfield  {journal} {\bibinfo  {journal} {J.
  Phys. Soc. Jpn.}\ }\textbf {\bibinfo {volume} {92}},\ \bibinfo {pages}
  {121007} (\bibinfo {year} {2023})}\BibitemShut {NoStop}%
\bibitem [{\citenamefont {Perry}(1994)}]{Perry_1994_PRSB}%
  \BibitemOpen
  \bibfield  {author} {\bibinfo {author} {\bibfnamefont {J.~N.}\ \bibnamefont
  {Perry}},\ }\href@noop {} {\bibfield  {journal} {\bibinfo  {journal} {Proc.
  Royal Soc. B}\ }\textbf {\bibinfo {volume} {257}},\ \bibinfo {pages} {221}
  (\bibinfo {year} {1994})}\BibitemShut {NoStop}%
\bibitem [{\citenamefont {Reuman}\ \emph {et~al.}(2017)\citenamefont {Reuman},
  \citenamefont {Zhaoa}, \citenamefont {Sheppard}, \citenamefont {Reid},\ and\
  \citenamefont {Cohen}}]{Reuman_etal_2017_PNAS}%
  \BibitemOpen
  \bibfield  {author} {\bibinfo {author} {\bibfnamefont {D.~C.}\ \bibnamefont
  {Reuman}}, \bibinfo {author} {\bibfnamefont {L.}~\bibnamefont {Zhaoa}},
  \bibinfo {author} {\bibfnamefont {L.~W.}\ \bibnamefont {Sheppard}}, \bibinfo
  {author} {\bibfnamefont {P.~C.}\ \bibnamefont {Reid}},\ and\ \bibinfo
  {author} {\bibfnamefont {J.~E.}\ \bibnamefont {Cohen}},\ }\href
  {https://doi.org/10.1073/pnas.1703593114} {\bibfield  {journal} {\bibinfo
  {journal} {Proc. Natl. Acad. Sci. U.S.A.}\ }\textbf {\bibinfo {volume}
  {114}},\ \bibinfo {pages} {6788} (\bibinfo {year} {2017})}\BibitemShut
  {NoStop}%
\bibitem [{\citenamefont {Giometto}\ \emph {et~al.}(2015)\citenamefont
  {Giometto}, \citenamefont {Formentin}, \citenamefont {Rinaldo}, \citenamefont
  {Cohen},\ and\ \citenamefont {Maritan}}]{Giometto_etal_2015_PNAS}%
  \BibitemOpen
  \bibfield  {author} {\bibinfo {author} {\bibfnamefont {A.}~\bibnamefont
  {Giometto}}, \bibinfo {author} {\bibfnamefont {M.}~\bibnamefont {Formentin}},
  \bibinfo {author} {\bibfnamefont {A.}~\bibnamefont {Rinaldo}}, \bibinfo
  {author} {\bibfnamefont {J.~E.}\ \bibnamefont {Cohen}},\ and\ \bibinfo
  {author} {\bibfnamefont {A.}~\bibnamefont {Maritan}},\ }\href
  {https://doi.org/10.1073/pnas.1505882112} {\bibfield  {journal} {\bibinfo
  {journal} {Proc. Natl. Acad. Sci. U.S.A.}\ }\textbf {\bibinfo {volume}
  {112}},\ \bibinfo {pages} {7755} (\bibinfo {year} {2015})}\BibitemShut
  {NoStop}%
\bibitem [{\citenamefont {Cohen}\ and\ \citenamefont
  {Xu}(2015)}]{Cohen_Xu_2015_PNAS}%
  \BibitemOpen
  \bibfield  {author} {\bibinfo {author} {\bibfnamefont {J.~E.}\ \bibnamefont
  {Cohen}}\ and\ \bibinfo {author} {\bibfnamefont {M.}~\bibnamefont {Xu}},\
  }\href {https://doi.org/10.1073/pnas.1503824112} {\bibfield  {journal}
  {\bibinfo  {journal} {Proc. Natl. Acad. Sci. U.S.A.}\ }\textbf {\bibinfo
  {volume} {112}},\ \bibinfo {pages} {7749} (\bibinfo {year}
  {2015})}\BibitemShut {NoStop}%
\bibitem [{\citenamefont {{F. Ballantyne IV}}(2005)}]{Ballantyne_2005_EER}%
  \BibitemOpen
  \bibfield  {author} {\bibinfo {author} {\bibnamefont {{F. Ballantyne IV}}},\
  }\href@noop {} {\bibfield  {journal} {\bibinfo  {journal} {Evol. Ecol. Res.}\
  }\textbf {\bibinfo {volume} {7}},\ \bibinfo {pages} {1213} (\bibinfo {year}
  {2005})}\BibitemShut {NoStop}%
\bibitem [{\citenamefont {{F. Ballantyne IV and A. J.
  Kerkhoff}}(2005)}]{Ballantyne_Kerkhoff_2005_JTB}%
  \BibitemOpen
  \bibfield  {author} {\bibinfo {author} {\bibnamefont {{F. Ballantyne IV and
  A. J. Kerkhoff}}},\ }\href {https://doi.org/10.1016/j.jtbi.2005.01.017}
  {\bibfield  {journal} {\bibinfo  {journal} {J. Theor. Biol.}\ }\textbf
  {\bibinfo {volume} {235}},\ \bibinfo {pages} {373} (\bibinfo {year}
  {2005})}\BibitemShut {NoStop}%
\bibitem [{\citenamefont {{F. Ballantyne IV and A. J.
  Kerkhoff}}(2007)}]{Ballantyne_Kerkhoff_2007_Oikos}%
  \BibitemOpen
  \bibfield  {author} {\bibinfo {author} {\bibnamefont {{F. Ballantyne IV and
  A. J. Kerkhoff}}},\ }\href {https://doi.org/10.1111/j.2006.0030-1299.15383.x}
  {\bibfield  {journal} {\bibinfo  {journal} {Oikos}\ }\textbf {\bibinfo
  {volume} {116}},\ \bibinfo {pages} {174} (\bibinfo {year}
  {2007})}\BibitemShut {NoStop}%
\bibitem [{\citenamefont {{A. J. Kerkhoff and F. Ballantyne
  IV}}(2003)}]{Kerkhoff_Ballantyne_2003_EL}%
  \BibitemOpen
  \bibfield  {author} {\bibinfo {author} {\bibnamefont {{A. J. Kerkhoff and F.
  Ballantyne IV}}},\ }\href {https://doi.org/10.1046/j.1461-0248.2003.00513.x}
  {\bibfield  {journal} {\bibinfo  {journal} {Ecol. Lett.}\ }\textbf {\bibinfo
  {volume} {6}},\ \bibinfo {pages} {850} (\bibinfo {year} {2003})}\BibitemShut
  {NoStop}%
\bibitem [{\citenamefont {Cohen}(2014{\natexlab{a}})}]{Cohen_2014_TE}%
  \BibitemOpen
  \bibfield  {author} {\bibinfo {author} {\bibfnamefont {J.~E.}\ \bibnamefont
  {Cohen}},\ }\href {https://doi.org/10.1007/s12080-013-0199-z} {\bibfield
  {journal} {\bibinfo  {journal} {Theor. Ecol.}\ }\textbf {\bibinfo {volume}
  {7}},\ \bibinfo {pages} {77} (\bibinfo {year}
  {2014}{\natexlab{a}})}\BibitemShut {NoStop}%
\bibitem [{\citenamefont {Cohen}(2014{\natexlab{b}})}]{Cohen_2014_TPB}%
  \BibitemOpen
  \bibfield  {author} {\bibinfo {author} {\bibfnamefont {J.~E.}\ \bibnamefont
  {Cohen}},\ }\href {https://doi.org/10.1016/j.tpb.2014.01.001} {\bibfield
  {journal} {\bibinfo  {journal} {Theor. Popul. Biol.}\ }\textbf {\bibinfo
  {volume} {93}},\ \bibinfo {pages} {30} (\bibinfo {year}
  {2014}{\natexlab{b}})}\BibitemShut {NoStop}%
\bibitem [{\citenamefont {Cohen}\ \emph {et~al.}(2013)\citenamefont {Cohen},
  \citenamefont {Xu},\ and\ \citenamefont {Schuster}}]{Cohen_etal_2013_PRSB}%
  \BibitemOpen
  \bibfield  {author} {\bibinfo {author} {\bibfnamefont {J.~E.}\ \bibnamefont
  {Cohen}}, \bibinfo {author} {\bibfnamefont {M.}~\bibnamefont {Xu}},\ and\
  \bibinfo {author} {\bibfnamefont {W.~S.}\ \bibnamefont {Schuster}},\
  }\href@noop {} {\bibfield  {journal} {\bibinfo  {journal} {Proc. Royal Soc.
  B}\ }\textbf {\bibinfo {volume} {280}} (\bibinfo {year} {2013})}\BibitemShut
  {NoStop}%
\bibitem [{\citenamefont {Zhao}\ \emph {et~al.}(2019)\citenamefont {Zhao},
  \citenamefont {Sheppard}, \citenamefont {Reid}, \citenamefont {Walter},\ and\
  \citenamefont {Reuman}}]{Zhao_etal_2019_JAE}%
  \BibitemOpen
  \bibfield  {author} {\bibinfo {author} {\bibfnamefont {L.}~\bibnamefont
  {Zhao}}, \bibinfo {author} {\bibfnamefont {L.~W.}\ \bibnamefont {Sheppard}},
  \bibinfo {author} {\bibfnamefont {P.~C.}\ \bibnamefont {Reid}}, \bibinfo
  {author} {\bibfnamefont {J.~A.}\ \bibnamefont {Walter}},\ and\ \bibinfo
  {author} {\bibfnamefont {D.~C.}\ \bibnamefont {Reuman}},\ }\href
  {https://doi.org/10.1111/1365-2656.12931} {\bibfield  {journal} {\bibinfo
  {journal} {J. Anim. Ecol.}\ }\textbf {\bibinfo {volume} {88}},\ \bibinfo
  {pages} {484} (\bibinfo {year} {2019})}\BibitemShut {NoStop}%
\bibitem [{\citenamefont {Cohen}(2013)}]{Cohen_2013_TPB}%
  \BibitemOpen
  \bibfield  {author} {\bibinfo {author} {\bibfnamefont {J.~E.}\ \bibnamefont
  {Cohen}},\ }\href {https://doi.org/10.1016/j.tpb.2013.04.002} {\bibfield
  {journal} {\bibinfo  {journal} {Theor. Popul. Biol.}\ }\textbf {\bibinfo
  {volume} {88}},\ \bibinfo {pages} {94} (\bibinfo {year} {2013})}\BibitemShut
  {NoStop}%
\bibitem [{\citenamefont {Cohen}\ and\ \citenamefont
  {Saitoh}(2016)}]{Cohen_Saitoh_2016_Ecology}%
  \BibitemOpen
  \bibfield  {author} {\bibinfo {author} {\bibfnamefont {J.~E.}\ \bibnamefont
  {Cohen}}\ and\ \bibinfo {author} {\bibfnamefont {T.}~\bibnamefont {Saitoh}},\
  }\href {https://doi.org/10.1002/ecy.1575} {\bibfield  {journal} {\bibinfo
  {journal} {Ecology}\ }\textbf {\bibinfo {volume} {97}},\ \bibinfo {pages}
  {3402} (\bibinfo {year} {2016})}\BibitemShut {NoStop}%
\bibitem [{\citenamefont {Saitoh}(2020)}]{Saitoh_2020_PE}%
  \BibitemOpen
  \bibfield  {author} {\bibinfo {author} {\bibfnamefont {T.}~\bibnamefont
  {Saitoh}},\ }\href {https://doi.org/10.1002/1438-390X.12051} {\bibfield
  {journal} {\bibinfo  {journal} {Popul. Ecol.}\ }\textbf {\bibinfo {volume}
  {62}},\ \bibinfo {pages} {300} (\bibinfo {year} {2020})}\BibitemShut
  {NoStop}%
\bibitem [{\citenamefont {Anderson}\ \emph {et~al.}(1982)\citenamefont
  {Anderson}, \citenamefont {Gordon}, \citenamefont {Crawley},\ and\
  \citenamefont {Hassel}}]{Anderson_etal_1982_Nature}%
  \BibitemOpen
  \bibfield  {author} {\bibinfo {author} {\bibfnamefont {R.~M.}\ \bibnamefont
  {Anderson}}, \bibinfo {author} {\bibfnamefont {D.~M.}\ \bibnamefont
  {Gordon}}, \bibinfo {author} {\bibfnamefont {M.~J.}\ \bibnamefont
  {Crawley}},\ and\ \bibinfo {author} {\bibfnamefont {M.~P.}\ \bibnamefont
  {Hassel}},\ }\href {https://doi.org/https://doi.org/10.1038/296245a0}
  {\bibfield  {journal} {\bibinfo  {journal} {Nature}\ }\textbf {\bibinfo
  {volume} {296}},\ \bibinfo {pages} {245} (\bibinfo {year}
  {1982})}\BibitemShut {NoStop}%
\bibitem [{\citenamefont {Kilpatrick}\ and\ \citenamefont
  {Ives}(2003)}]{Kilpatrick_Ives_2003_Nature}%
  \BibitemOpen
  \bibfield  {author} {\bibinfo {author} {\bibfnamefont {A.~M.}\ \bibnamefont
  {Kilpatrick}}\ and\ \bibinfo {author} {\bibfnamefont {A.~R.}\ \bibnamefont
  {Ives}},\ }\href {https://doi.org/10.1038/nature01471} {\bibfield  {journal}
  {\bibinfo  {journal} {Nature}\ }\textbf {\bibinfo {volume} {422}},\ \bibinfo
  {pages} {65} (\bibinfo {year} {2003})}\BibitemShut {NoStop}%
\bibitem [{\citenamefont {Carpenter}\ \emph {et~al.}(2023)\citenamefont
  {Carpenter}, \citenamefont {Callens}, \citenamefont {Brown}, \citenamefont
  {Cohen},\ and\ \citenamefont {Webb}}]{Carpenter_etal_2023_TPB}%
  \BibitemOpen
  \bibfield  {author} {\bibinfo {author} {\bibfnamefont {S.}~\bibnamefont
  {Carpenter}}, \bibinfo {author} {\bibfnamefont {S.}~\bibnamefont {Callens}},
  \bibinfo {author} {\bibfnamefont {C.}~\bibnamefont {Brown}}, \bibinfo
  {author} {\bibfnamefont {J.~E.}\ \bibnamefont {Cohen}},\ and\ \bibinfo
  {author} {\bibfnamefont {B.~Z.}\ \bibnamefont {Webb}},\ }\href
  {https://doi.org/10.1016/j.tpb.2023.10.002} {\bibfield  {journal} {\bibinfo
  {journal} {Theor. Popul. Biol.}\ }\textbf {\bibinfo {volume} {154}},\
  \bibinfo {pages} {118} (\bibinfo {year} {2023})}\BibitemShut {NoStop}%
\bibitem [{\citenamefont {Taylor}\ and\ \citenamefont
  {Woiwod}(1982)}]{Taylor_Woiwod_1982_JAE}%
  \BibitemOpen
  \bibfield  {author} {\bibinfo {author} {\bibfnamefont {L.~R.}\ \bibnamefont
  {Taylor}}\ and\ \bibinfo {author} {\bibfnamefont {I.~P.}\ \bibnamefont
  {Woiwod}},\ }\href@noop {} {\bibfield  {journal} {\bibinfo  {journal} {J.
  Anim. Ecol.}\ }\textbf {\bibinfo {volume} {51}},\ \bibinfo {pages} {879}
  (\bibinfo {year} {1982})}\BibitemShut {NoStop}%
\bibitem [{\citenamefont {Hanski}(1987)}]{Hanski_1987_Oikos}%
  \BibitemOpen
  \bibfield  {author} {\bibinfo {author} {\bibfnamefont {I.}~\bibnamefont
  {Hanski}},\ }\href {https://www.jstor.org/stable/3565413?seq=1&cid=pdf-}
  {\bibfield  {journal} {\bibinfo  {journal} {Oikos}\ }\textbf {\bibinfo
  {volume} {50}},\ \bibinfo {pages} {148} (\bibinfo {year} {1987})}\BibitemShut
  {NoStop}%
\bibitem [{\citenamefont {Blasius}\ \emph {et~al.}(1999)\citenamefont
  {Blasius}, \citenamefont {Huppert},\ and\ \citenamefont
  {Stone}}]{Blasius_etal_1999_Nature}%
  \BibitemOpen
  \bibfield  {author} {\bibinfo {author} {\bibfnamefont {B.}~\bibnamefont
  {Blasius}}, \bibinfo {author} {\bibfnamefont {A.}~\bibnamefont {Huppert}},\
  and\ \bibinfo {author} {\bibfnamefont {L.}~\bibnamefont {Stone}},\
  }\href@noop {} {\bibfield  {journal} {\bibinfo  {journal} {Nature}\ }\textbf
  {\bibinfo {volume} {399}},\ \bibinfo {pages} {354} (\bibinfo {year}
  {1999})}\BibitemShut {NoStop}%
\bibitem [{\citenamefont {Blasius}\ and\ \citenamefont
  {Stone}(2000)}]{Blasius_Stone_2000_IJBC}%
  \BibitemOpen
  \bibfield  {author} {\bibinfo {author} {\bibfnamefont {B.}~\bibnamefont
  {Blasius}}\ and\ \bibinfo {author} {\bibfnamefont {L.}~\bibnamefont
  {Stone}},\ }\href@noop {} {\bibfield  {journal} {\bibinfo  {journal} {Int. J.
  Bifurcat. Chaos}\ }\textbf {\bibinfo {volume} {10}} (\bibinfo {year}
  {2000})}\BibitemShut {NoStop}%
\bibitem [{\citenamefont {Montbri^^c3^^b3}\ and\ \citenamefont
  {Blasius}(2003)}]{Montbrio_Blasius_2003_Chaos}%
  \BibitemOpen
  \bibfield  {author} {\bibinfo {author} {\bibfnamefont {E.}~\bibnamefont
  {Montbri^^c3^^b3}}\ and\ \bibinfo {author} {\bibfnamefont {B.}~\bibnamefont
  {Blasius}},\ }\href {https://doi.org/10.1063/1.1525170} {\bibfield  {journal}
  {\bibinfo  {journal} {Chaos}\ }\textbf {\bibinfo {volume} {13}},\ \bibinfo
  {pages} {291} (\bibinfo {year} {2003})}\BibitemShut {NoStop}%
\bibitem [{\citenamefont {Rosenblum}\ \emph {et~al.}(1996)\citenamefont
  {Rosenblum}, \citenamefont {Pikovsky},\ and\ \citenamefont
  {Kurths}}]{Rosenblum_etal_1996_PRL}%
  \BibitemOpen
  \bibfield  {author} {\bibinfo {author} {\bibfnamefont {M.~G.}\ \bibnamefont
  {Rosenblum}}, \bibinfo {author} {\bibfnamefont {A.~S.}\ \bibnamefont
  {Pikovsky}},\ and\ \bibinfo {author} {\bibfnamefont {J.}~\bibnamefont
  {Kurths}},\ }\href@noop {} {\bibfield  {journal} {\bibinfo  {journal} {Phys.
  Rev. Lett.}\ }\textbf {\bibinfo {volume} {76}},\ \bibinfo {pages} {1804}
  (\bibinfo {year} {1996})}\BibitemShut {NoStop}%
\bibitem [{\citenamefont {Rosenblum}\ \emph {et~al.}(1997)\citenamefont
  {Rosenblum}, \citenamefont {Pikovsky},\ and\ \citenamefont
  {Kurths}}]{Rosenblum_etal_1997_PRL}%
  \BibitemOpen
  \bibfield  {author} {\bibinfo {author} {\bibfnamefont {M.~G.}\ \bibnamefont
  {Rosenblum}}, \bibinfo {author} {\bibfnamefont {A.~S.}\ \bibnamefont
  {Pikovsky}},\ and\ \bibinfo {author} {\bibfnamefont {J.}~\bibnamefont
  {Kurths}},\ }\href@noop {} {\bibfield  {journal} {\bibinfo  {journal} {Phys.
  Rev. Lett.}\ }\textbf {\bibinfo {volume} {78}},\ \bibinfo {pages} {4193}
  (\bibinfo {year} {1997})}\BibitemShut {NoStop}%
\bibitem [{\citenamefont {Sakaguchi}(2000)}]{Sakaguchi_2000_PRE}%
  \BibitemOpen
  \bibfield  {author} {\bibinfo {author} {\bibfnamefont {H.}~\bibnamefont
  {Sakaguchi}},\ }\href {https://doi.org/10.1103/PhysRevE.61.7212} {\bibfield
  {journal} {\bibinfo  {journal} {Phys. Rev. E}\ }\textbf {\bibinfo {volume}
  {61}},\ \bibinfo {pages} {7212} (\bibinfo {year} {2000})}\BibitemShut
  {NoStop}%
\bibitem [{\citenamefont {Pikovsky}\ \emph {et~al.}(1996)\citenamefont
  {Pikovsky}, \citenamefont {Rosenblum},\ and\ \citenamefont
  {Kurths}}]{Pikovsky_etal_1996_EPL}%
  \BibitemOpen
  \bibfield  {author} {\bibinfo {author} {\bibfnamefont {A.~S.}\ \bibnamefont
  {Pikovsky}}, \bibinfo {author} {\bibfnamefont {M.~G.}\ \bibnamefont
  {Rosenblum}},\ and\ \bibinfo {author} {\bibfnamefont {J.}~\bibnamefont
  {Kurths}},\ }\href {https://doi.org/10.1209/epl/i1996-00433-3} {\bibfield
  {journal} {\bibinfo  {journal} {Europhys. Lett.}\ }\textbf {\bibinfo {volume}
  {34}},\ \bibinfo {pages} {165} (\bibinfo {year} {1996})}\BibitemShut
  {NoStop}%
\bibitem [{\citenamefont {Osipov}\ \emph {et~al.}(1997)\citenamefont {Osipov},
  \citenamefont {Pikovsky}, \citenamefont {Rosenblum},\ and\ \citenamefont
  {Kurths}}]{Osipov_etal_1997_PRE}%
  \BibitemOpen
  \bibfield  {author} {\bibinfo {author} {\bibfnamefont {G.~V.}\ \bibnamefont
  {Osipov}}, \bibinfo {author} {\bibfnamefont {A.~S.}\ \bibnamefont
  {Pikovsky}}, \bibinfo {author} {\bibfnamefont {M.~G.}\ \bibnamefont
  {Rosenblum}},\ and\ \bibinfo {author} {\bibfnamefont {J.~R.}\ \bibnamefont
  {Kurths}},\ }\href@noop {} {\bibinfo {title} {Phase synchronization effects
  in a lattice of nonidentical r^^c3^^b6 ssler oscillators}} (\bibinfo {year}
  {1997})\BibitemShut {NoStop}%
\bibitem [{\citenamefont {Hanski}\ and\ \citenamefont
  {Woiwod}(1993)}]{Hanski_Woiwod_1993_JAE}%
  \BibitemOpen
  \bibfield  {author} {\bibinfo {author} {\bibfnamefont {I.}~\bibnamefont
  {Hanski}}\ and\ \bibinfo {author} {\bibfnamefont {I.~P.}\ \bibnamefont
  {Woiwod}},\ }\href {https://about.jstor.org/terms} {\bibfield  {journal}
  {\bibinfo  {journal} {J. Anim. Ecol.}\ }\textbf {\bibinfo {volume} {62}},\
  \bibinfo {pages} {656} (\bibinfo {year} {1993})}\BibitemShut {NoStop}%
\bibitem [{\citenamefont {Liebhold}\ \emph {et~al.}(2004)\citenamefont
  {Liebhold}, \citenamefont {Koenig},\ and\ \citenamefont
  {Bj^^c3^^b8rnstad}}]{Liebhold_etal_2004_review}%
  \BibitemOpen
  \bibfield  {author} {\bibinfo {author} {\bibfnamefont {A.}~\bibnamefont
  {Liebhold}}, \bibinfo {author} {\bibfnamefont {W.~D.}\ \bibnamefont
  {Koenig}},\ and\ \bibinfo {author} {\bibfnamefont {O.~N.}\ \bibnamefont
  {Bj^^c3^^b8rnstad}},\ }\href
  {https://doi.org/10.2307/annurev.ecolsys.34.011802.30000018} {\bibfield
  {journal} {\bibinfo  {journal} {Annu. Rev. Ecol. Evol. Sys}\ }\textbf
  {\bibinfo {volume} {35}},\ \bibinfo {pages} {467} (\bibinfo {year}
  {2004})}\BibitemShut {NoStop}%
\bibitem [{\citenamefont {Petri}\ \emph {et~al.}(2013)\citenamefont {Petri},
  \citenamefont {Expert}, \citenamefont {Jensen},\ and\ \citenamefont
  {Polak}}]{Petri_etal_2013_SR}%
  \BibitemOpen
  \bibfield  {author} {\bibinfo {author} {\bibfnamefont {G.}~\bibnamefont
  {Petri}}, \bibinfo {author} {\bibfnamefont {P.}~\bibnamefont {Expert}},
  \bibinfo {author} {\bibfnamefont {H.~J.}\ \bibnamefont {Jensen}},\ and\
  \bibinfo {author} {\bibfnamefont {J.~W.}\ \bibnamefont {Polak}},\ }\href@noop
  {} {\bibfield  {journal} {\bibinfo  {journal} {Sci. Rep.}\ }\textbf {\bibinfo
  {volume} {3}} (\bibinfo {year} {2013})}\BibitemShut {NoStop}%
\bibitem [{\citenamefont {Mainieri}\ and\ \citenamefont
  {Rehacek}(1999)}]{Mainieri_Rehacek_1999_PRL}%
  \BibitemOpen
  \bibfield  {author} {\bibinfo {author} {\bibfnamefont {R.}~\bibnamefont
  {Mainieri}}\ and\ \bibinfo {author} {\bibfnamefont {J.}~\bibnamefont
  {Rehacek}},\ }\href@noop {} {\bibfield  {journal} {\bibinfo  {journal} {Phys.
  Rev. Lett.}\ }\textbf {\bibinfo {volume} {82}},\ \bibinfo {pages} {3042}
  (\bibinfo {year} {1999})}\BibitemShut {NoStop}%
\bibitem [{\citenamefont {Kano}\ and\ \citenamefont
  {Umeno}(2022)}]{Kano_Umeno_2022_Chaos}%
  \BibitemOpen
  \bibfield  {author} {\bibinfo {author} {\bibfnamefont {T.}~\bibnamefont
  {Kano}}\ and\ \bibinfo {author} {\bibfnamefont {K.}~\bibnamefont {Umeno}},\
  }\href {https://doi.org/10.1063/5.0100897} {\bibfield  {journal} {\bibinfo
  {journal} {Chaos}\ }\textbf {\bibinfo {volume} {32}},\ \bibinfo {pages}
  {113137} (\bibinfo {year} {2022})}\BibitemShut {NoStop}%
\end{thebibliography}%

\end{document}